\documentclass[10pt,conference]{IEEEtran}
\IEEEoverridecommandlockouts
\usepackage{cite}
\usepackage{amsmath,amssymb,amsfonts}
\usepackage{algorithmic}
\usepackage{graphicx}
\usepackage{textcomp}
\usepackage{xcolor}

\usepackage{algorithmic}
\usepackage{algorithm}
\renewcommand{\algorithmicrequire}{\textbf{Input:}} 
\renewcommand{\algorithmicensure}{\textbf{Output:}} 

\usepackage{changepage}

\usepackage{amsmath}

\usepackage{threeparttable}

\usepackage{multirow} 

\usepackage{array} 

\usepackage{booktabs} 

\usepackage{amssymb}
\newcommand{\RomanNumeral}[1]{\text{\MakeUppercase{\romannumeral#1}}}

\makeatletter
\let\NAT@parse\undefined
\makeatother
\usepackage{hyperref}  

\def\BibTeX{{\rm B\kern-.05em{\sc i\kern-.025em b}\kern-.08em
    T\kern-.1667em\lower.7ex\hbox{E}\kern-.125emX}}

\DeclareRobustCommand*{\IEEEauthorrefmark}[1]{%
    \raisebox{0pt}[0pt][0pt]{\textsuperscript{\footnotesize\ensuremath{#1}}}}

\begin{document}

\title{A Microservices Identification Method Based on Spectral Clustering for Industrial Legacy Systems}

\author{
\IEEEauthorblockN{
Teng Zhong\IEEEauthorrefmark{1},
Yinglei Teng\IEEEauthorrefmark{1}\IEEEauthorrefmark{*}, \emph{Senior Member}, \emph{IEEE}, 
Shijun Ma\IEEEauthorrefmark{1},
Jiaxuan Chen\IEEEauthorrefmark{1}, and
Sicong Yu\IEEEauthorrefmark{2}}
\IEEEauthorblockA{\IEEEauthorrefmark{1}Beijing Key Laboratory of Work Safety Intelligent Monitoring,\\
Beijing University of Post and Telecommunications, Beijing, China}
\IEEEauthorblockA{\IEEEauthorrefmark{2}Technology and Standards Research Institute, \\
China Academy of Information and Communications
Technology, Beijing, China}
\IEEEauthorblockA{Email: \{lilytengtt\IEEEauthorrefmark{*}, zhongteng, chenjiaxuan, mashijun\}@bupt.edu.cn}
}

\maketitle
\begin{abstract}
The advent of Industrial Internet of Things (IIoT) has imposed more stringent requirements on industrial software in terms of communication delay, scalability, and maintainability. Microservice architecture (MSA), a novel software architecture that has emerged from cloud computing and DevOps, presents itself as the most promising solution due to its independently deployable and loosely coupled nature. Currently, practitioners are inclined to migrate industrial legacy systems to MSA, despite numerous challenges it presents. In this paper, we propose an automated microservice decomposition method for extracting microservice candidates based on spectral graph theory to address the problems associated with manual extraction, which is time-consuming, labor-intensive, and highly subjective. The method is divided into three steps. Firstly, static and dynamic analysis tools are employed to extract dependency information of the legacy system. Subsequently, information is transformed into a graph structure that captures inter-class structure and performance relationships in legacy systems. Finally, graph-based clustering algorithm is utilized to identify potential microservice candidates that conform to the principles of high cohesion and low coupling. Comparative experiments with state-of-the-art methods demonstrate the significant advantages of our proposed method in terms of performance metrics. Moreover, Practice show that our method can yield favorable results even without the involvement of domain experts.
\end{abstract}


\begin{figure*}[htbp]
  \centering
  \includegraphics[width = 1.0\textwidth]{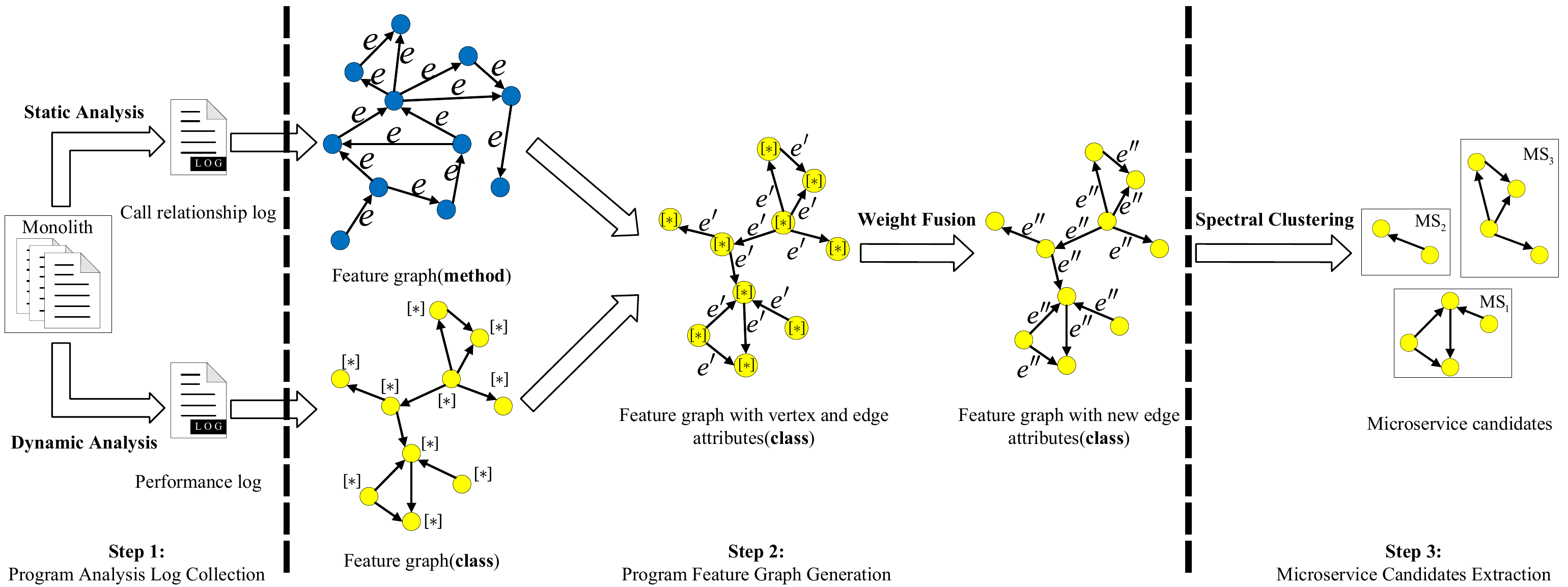}
  \caption{The overview of our spectral clustering microservice candidate extraction method}
  \label{fig: overview}
\end{figure*}

\section{Introduction}
With the deep integration of IT and OT, the industry is reconsidering the challenges faced by the Industrial Internet of Things (IIoT). The traditional industrial software, which follows a monolithic architecture (MA), falls short in meeting the new requirements of the industrial network regarding communication delay and reconfigurability due to its poor maintainability and scalability. In recent years, the advancement of cloud computing technology has brought attention to an innovative software architecture named Microservice Architecture (MSA)\cite{newman2021building}. Studies and practical experience have demonstrated that the MSA exhibits characteristics such as small and autonomous services, a lightweight communication protocol, and compatibility with various technology stacks. Consequently, it has emerged as a promising approach for refactoring industrial software.

Despite companies having recognized the diverse benefits offered by MSA, there remain many industrial legacy systems that continue to operate under MA. The reasons are manifold, the most significant one being that legacy systems have been operating under the MA for an extended period. Personnel within the company perhaps have extensively expressed dissatisfaction with this system, citing critical issues like redundant code, intricate logic, and even a lack of design documentation. However, developing a distributed software system from scratch, with the aim of emulating the functionalities of the original system, would be impractical with the aspect of the required investment and human effort involved. Therefore, most refactoring methods involve starting from the legacy code base of the industrial legacy systems and extracting a set of software artifacts based on MSA design principles before deploying them as microservices\cite{adjoyan2014service}. The extraction process is referred to as microservice candidates identification.

In recent years, how to extract microservice candidates in an elegant manner has been a prominent issue that has garnered attention from both industry and academia. Currently, the prevalent approach in the industry involves manually identifying candidates based on predefined rules. However, such methods heavily rely on the expertise of practitioners and are often limited to specific legacy systems, lacking universality. On the other hand, the academic community tends to utilize various technologies to extract relevant characteristic information from the system to be migrated. The extracted information is then used as input for microservice candidates identification algorithms. This type methodology has the advantage of being adaptable to various type of legacy systems. Nevertheless, its effectiveness is contingent upon the richness of feature information and the algorithm employed.


The use of microservice candidates extraction algorithms can be traced back to \cite{gysel2016service}. Gysel et al. manually analyzed the legacy system and transformed it into system specification artifacts. The generated artifacts were then inputted into a graph clustering algorithm to generate microservice candidates that satisfy their proposed coupling criteria. Another approach was taken by Zhang et al.\cite{zhang2020automated}, who collected functional logs and non-functional logs from the legacy system using dynamic analysis tools. They employed a genetic algorithm with three objectives to identify the optimal result for service identification. Agarwal et al.\cite{agarwal2021monolith} conducted a search for seed classes within legacy systems by utilizing formal concept analysis. They then used the density clustering algorithm to extract microservice candidates, as well as unassigned and non-functional groups. Desai et al.\cite{desai2021graph} utilized static program analysis tools to obtain entry points and structural dependencies of legacy programs, which were transformed into graph structures with vertex attributes. They then used graph convolution neural networks and a step-by-step training method to modify the loss function and extract microservice candidates. Li et al.\cite{li2022microservice} performed microservice candidate extraction using knowledge graphs based on AKF principles. Zaragoza et al.\cite{zaragoza2022leveraging} extracted microservice candidates by analyzing the layer architectures of the legacy system. Although these methods employ algorithms for extracting microservice candidates, they still require manual analysis or the involvement of domain experts to extract feature information. These human interventions remain time-consuming, labor-intensive, and dependent on the expertise of operators. While certain approaches have utilized automated feature extraction tools to obtain feature information, a critical issue is that the incomprehensiveness of the extracted information still exists.

To address the time-consuming and laborious issue of manually extracting microservice candidates, we propose an automated microservice candidate extraction method based on the principle of high cohesion and low coupling. Our approach encompasses a comprehensive set of operational procedures, starting from the legacy system code base. By employing static and dynamic program analysis tools, we automatically extract extensive feature information from the legacy system. During the phase of microservice candidate identification, we model corresponding optimization goals, perform mathematical derivations, and utilize a machine learning algorithm based on spectral graph theory to automatically identify potential microservice candidates. To the best of our knowledge, our team is the first to utilize this theory for automatic microservice candidate identification. Experimental results demonstrate that our proposed method achieves superior performance metrics compared to current state-of-the-art approaches.

The remainder of the paper is structured as follows: Section \RomanNumeral{2} details our proposed method. Section \RomanNumeral{3} shows the experimental results. Section \RomanNumeral{4} concludes the paper.

\section{Methodology}
In this section, we propose a spectral clustering microservice candidate extraction method based on both static and dynamic program analysis. As illustrated in Fig. \ref{fig: overview},the whole process includes three main steps. In step \RomanNumeral{1}, call relationship logs and performance logs are acquired from the legacy system using both static and dynamic analysis tools. In step \RomanNumeral{2}, we represent the legacy system as a graph with attributes for both edge and vertex, thereby exploiting the inherent graph structure of software systems. These attributes are then merged to create a new graph structure that is suitable for the content-insensitive clustering algorithm. In step \RomanNumeral{3}, the spectral clustering algorithm is utilized to generate a set of microservice candidates that adhere to the software design principle of high cohesion and loose coupling. We will provide detailed illustrations to these three steps in the rest of section.

\begin{figure*}[tbp]
  \centering
  \includegraphics[width = 0.75\textwidth]{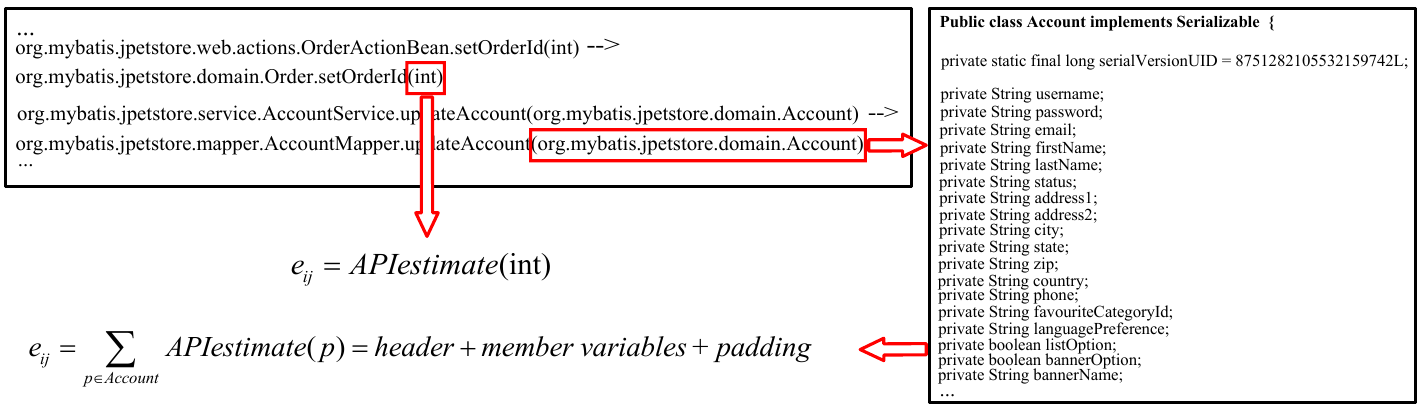}
  \caption{Calculation rule of remote call overhead (take jpetstore-6 as an example).}
  \label{fig: static log}
\end{figure*}

\begin{figure}[tbp]
\centerline{\includegraphics[width = 0.75\columnwidth]{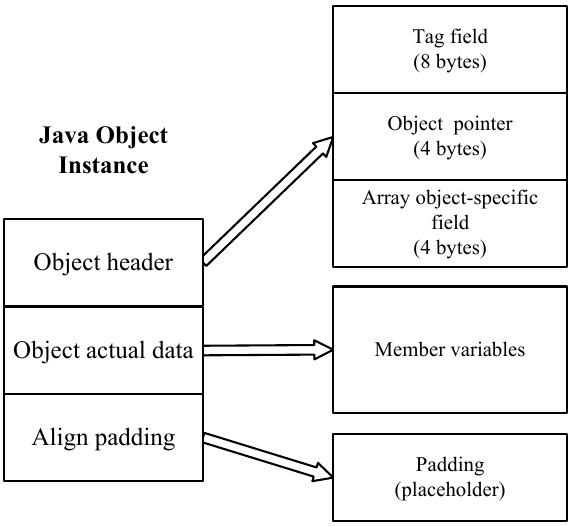}}
\caption{Java object instance diagram under 64-bit operating system.}
\label{fig: java object}
\end{figure}

\subsection{Program Analysis Log Collection}
To exemplify our approach, we utilize an open-source Java web program jpetstore-6\footnote{https://github.com/mybatis/jpetstore-6} as demo. The static analysis tool we first employed to gather structural dependencies within the legacy system. Using the system source code as input, this tool generates a log that includes method call pairs and input parameters. We collated it to call relationship log $L_{c}$, as portrayed in Eq. \ref{eq:equation1}. Each line of the call relationship log corresponds to a six-tuple ${{\bf{l}}_{\bf{c}}}$, encompassing the caller and callee methods, the corresponding Java classes, and the input parameters of the method definition.

\begin{equation}
{{\bf{l}}_{\bf{c}}} \in {L_c},\quad {{\bf{l}}_{\bf{c}}} = \{ {m_i},{m_j},{c_i},{c_j},{p_i},{p_j}\}.
\label{eq:equation1}
\end{equation}

In addition to capturing structural dependencies through static analysis, dynamic analysis tool was adopted to capture the runtime characteristics of the legacy system. By utilizing the stress testing tool Jmeter\footnote{https://github.com/apache/jmeter} and the performance monitoring tool Jvisualvm\footnote{https://github.com/oracle/visualvm} in conjunction, we can acquire the runtime characteristics of software artifacts within the legacy system. Inspired by \cite{2019Trans}, we analyze the business functions of the legacy system to design comprehensive test scenarios that contains as many functions as possible. The scenarios are then executed for functional testing, with the operations being recorded into thread groups using Jmeter's script recording function, allowing for repeated execution. During the execution of thread groups, Jvisualvm monitors the performance information of the software artifacts involved in the functional testing. We transform this information into the performance log $L_{p}$ as 

\begin{equation}
{{\bf{l}}_{\bf{p}}} \in {L_p},\quad {{\bf{l}}_{\bf{p}}} = \{ {c_i},{t_i},{r_i}\},
\label{eq:equation2}
\end{equation}
where $c_{i}$ represents the corresponding class identifier, $t_{i}$ denotes the CPU runtime of $c_{i}$ during functional testing, and $r_{i}$ represents the average retained memory size occupied by $c_{i}$ during functional testing.

\begin{table}[tbp]
\caption{The memory size occupied by basic variables.}
\renewcommand{\arraystretch}{1.25} 
\begin{center}
\scalebox{0.9}
{
\begin{threeparttable}
\begin{tabular}{|c|c|c|c|c|c|c|c|c|}
\hline
\textbf{Data type} & byte & short & int & long & char & float & double & boolean \\
\hline
\textbf{Size(byte)} & 1 & 2 & 4 & 8 & 2 & 4 & 8 & 4{*} \\
\hline
\end{tabular}

\begin{tablenotes}
    \footnotesize
    \item[*] When a boolean variable is declared alone, the JVM considers it as an $int$ type, occupying 4 bytes. If it's declared within an array, the JVM considers it as a $byte$ type, occupying 1 byte.
\end{tablenotes}
\end{threeparttable}
}
\label{tab1}
\end{center}
\end{table}

\subsection{Program Feature Graph Generation}
Through analysis of the static information from $L_{c}$, we establish method-level structural dependencies from the legacy system and represent it as a weighted directed graph. The edge weights are defined as the communication overhead caused by remote calls when methods are deployed in different services in the MSA. We quantify the edge weights based on the input parameters of callee $p_j$, as depicted in Eq. \ref{eq:equation3}. To calculate the overhead of each parameter $p$ in $p_j$, we define the APIestimate function. Its calculation rule we refer to the memory size occupied by various types of variables in the Java virtual machine (JVM)\cite{jendrock2014java}. The memory sizes of basic variable types are outlined in Tab. \ref{tab1}. For reference variable types in object-oriented languages, we refer to the structure diagram of Java object instances in the HotSpot JVM, as illustrated in Fig. \ref{fig: java object}. In 64-bit operating system, a Java object instance comprises three components: the object header, the object's actual data, and alignment padding. The object header, which consists of a tag field, an object pointer, and an array object-specific field, occupies 12 to 16 bytes. The memory size occupied by the object's actual data depends on the member variables present in the object instance. The purpose of alignment padding is to insert placeholders, ensuring that the memory size occupied by the entire Java object instance is always a multiple of 8 bytes. We demonstrate this rule using a portion of the call relationship log extracted from jpetstore-6, as depicted in Fig. \ref{fig: static log}.


\begin{equation}
{e_{ij}} = \sum\limits_{p \in {p_j}} {\text{APIestimate}(p)} + 1.
\label{eq:equation3}
\end{equation}

After collecting $L_{c}$, we represent the legacy system as a directed graph. In this graph, the methods serve as vertices, the structural dependencies serve as edges, and the communication overhead caused by remote calls serves as edge weights. This data structure contains the structural information of the legacy system, which can be used as input for the subsequent algorithm to extract microservice candidates. However, there are two important considerations: First, when migrating a legacy system to MSA, architects should avoid making internal changes to the original software artifacts of the legacy system. Second, generating microservice candidates at the method level would require significant time and labor costs for software engineers during the actual deployment. To address these concerns, we increase the granularity of the program feature graph to the class level. The rules for increasing granularity are shown in Eq. \ref{eq:equation4}.


\begin{equation}
{e'_{ij}} = \sum\limits_{{m_i \in c_i}, {m_j \in c_j}, i \neq j} {e_{ij}}.
\label{eq:equation4}
\end{equation}
We then enhance the class-level feature graph using the performance log $L_{p}$. Specifically, we add the CPU runtime $t_{i}$ and the average retained memory size $r_{i}$ as attributes to the vertex corresponding to $c_{i}$. This process results in a weighted graph structure with both edge attributes and node attributes. However, the subsequent algorithm we utilize is content-insensitive, meaning that if we directly use the class-level graph as input, the algorithm will only extract the edge attributes. In order to address this, we employ the skill proposed by \cite{2013CODICIL} to fuse the edge and node attributes of the graph and create a new graph structure that is suitable for content-insensitive algorithms, expressed as

\begin{equation}
{e''_{ij}} = {e'_{ij}} \times ({t_j} + {r_j} + 1).
\label{eq:equation5}
\end{equation}
Without deleting any vertices from the original graph, we fuse the edge and node attributes to generate new edge weights for the graph. Ultimately, we obtain a class-level fusion graph with $e''_{ij}$ as the edge weight. We represent this graph structure as Equation \ref{eq:equation6}.

\begin{equation}
G = (V, E), \quad V = \{{c_i}\}, \quad E = \{{e''_{ij}}\}.
\label{eq:equation6}
\end{equation}

\subsection{Microservice Candidates Extraction}
In this step, the process of extracting microservice candidates from a legacy system is modeled as a min-cut problem in the program feature graph. The objective is to generate connected components that adhere to the principles of high cohesion and low coupling. The optimization goal is depicted: 

\begin{equation}
\begin{aligned}
&\text{min Cut}(A_1, A_2, \dots, A_K) = \frac{1}{2} \sum_{k=1}^K W(A_k, \overline{A_k}), \\
&\text{s.t.} 
\begin{cases}
    G = (E, V), \quad E = [{e'}_{ij}], \quad V = \{v_1, v_2, \dots, v_N\}, \\
    A_k = \{v_i \mid v_i \in V\}, \quad V = \bigcup_{k=1}^K A_k, \\
    \forall i,j \in \{1,2,\dots,N\}, \quad A_i \cap A_j \neq \emptyset, \\
    1 \leq i \leq N, \quad 1 \leq k \leq K.
\end{cases}
\end{aligned}
\label{eq:equation7}
\end{equation}
In Eq. \ref{eq:equation7}, $A_{k}$ is the connected component that represents a microservice candidate, and ${W({A_k},\overline {{A_k}} )}$ is the cut of the connected component and its complement.The entire optimization objective is transformed into Eq. \ref{eq:equation8} through mathematical derivation and relaxation processing.

\begin{equation}
\hat{Y} = \underbrace{\operatorname*{arg\,min}_{Y \in \mathbb{R}^{N \times K}}}_\text{subject to} \operatorname*{tr}\left(Y^T L Y\right) \quad \text{s.t.} \quad Y^T Y = I,
\label{eq:equation8}
\end{equation}
where $Y$ is an indicator matrix used to signify the result of graph vertex division, and $L$ is a Laplacian matrix that stores the detail information of the graph's edge weight matrix, $W$. The solvability of the objective function in polynomial time is proven by the Rayleigh-Ritz theorem. Moreover, the objective function is consistent with the objective function of the spectral clustering algorithm. Therefore, we employ the graph-based spectral clustering algorithm\cite{von2007tutorial} to extract microservice candidates.

This algorithm offers several advantages compared to traditional prototype clustering algorithms: 1) Spectral clustering can naturally handle the min-cut problem in graph theory when the affinity matrix coincides with the graph's adjacency matrix, making it more suitable for datasets with connectivity, such as legacy systems. 2) As the core principle of spectral clustering involves using Laplacian feature mapping to reduce the dimension of samples, only an affinity matrix is needed as input to generate clustering results. 3) This algorithm also executes faster than prototype clustering algorithms. The pseudocode for our method is presented in Algorithm \ref{algorithm1}, where $K$ is the expected number of microservice candidates to be extracted, $C$ is the label set of the legacy program class files, $M_{s}$ is the set of microservice candidates, and $W_m$, $W_c$, $W_c^{'}$ are adjacency matrices of program feature graphs.




\begin{figure}[htbp]
    \label{algorithm1}
    \renewcommand{\algorithmicrequire}{\textbf{Input:}}
    \renewcommand{\algorithmicensure}{\textbf{Output:}}
    \begin{algorithm}[H]
        \caption{Microservice Candidates Extraction Algorithm}
        \begin{algorithmic}[1]
            \REQUIRE Call relationship log $L_c$, Performance log $L_p$, Number of microservice candidates $k$. 
            \ENSURE Microservice candidates set $M_s$.  
            
            \STATE // Obtain the similarity matrix for clustering
            \STATE Sort out the method-level adjacency matrix $W_m$ from $L_c$.
            \STATE Increase the granularity of $W_m$ to obtain the class-level adjacency matrix $W_c$ according to Eq. \ref{eq:equation4}.
            \STATE Obtain the adjacency matrix $W_c^{'}$ of the content-insensitive graph structure according to $L_p$ and Eq. \ref{eq:equation5}.
            
            \STATE // The procedure of Spectral Clustering
            \STATE Using $W_c^{'}$ to compute the unnormalized Laplacian $L$.
            \STATE Compute the first $k$ eigenvectors $u_1, \ldots, u_k$ of $L$.
            \STATE Let $U \in {\mathbb{R}^{n \times k}}$ be the matrix containing the vectors $u_1, \ldots, u_k$ as columns.       
            \FOR{$i = 1, \ldots, n$}
                \STATE Let ${y^i} \in {\mathbb{R}^k}$ be the vector corresponding to the $i$-th row of $U$.
            \ENDFOR
            \STATE Cluster the points ${\left( {{y_i}} \right)_{i = 1, \ldots, n}}$ in $\mathbb{R}^k$ using the K-Means algorithm into Microservice candidates set $M_s$.
        \end{algorithmic}
    \end{algorithm}
\end{figure}

\section{Experiment and Discussion}
In this section, we first introduce two questions inspired by our proposed method. 

\begin{adjustwidth}{0em}{0em} 
\hspace*{2em}\textit{\textbf{RQ1:} Will a content-insensitive graph structure improve the clustering results of our proposed method?}
\end{adjustwidth}

\begin{adjustwidth}{0em}{0em} 
\hspace*{2em}\textit{\textbf{RQ2:} Does our microservice candidate extraction method outperform the baselines in terms of performance?}
\end{adjustwidth}
The remainder of this section is organized as follows: we first introduce the evaluation metric used to assess the cohesion and coupling degrees of the microservice candidate set. Next, we provide an overview of the legacy systems used in the extraction of microservice candidates. Finally, we conduct relevant experiments to answer the raised questions and demonstrate the superiority of our proposed method.

\subsection{Metric for Assessing Cohesion and Coupling}
We utilize the modularity quality (MQ), introduced by Mancoridis et al. \cite{mancoridis1998using}, to evaluate the effectiveness of microservice candidate extraction. MQ is widely used for evaluating graph partitioning result, as shown in Eq. \ref{eq:equation9}. Specifically, it consists of two parts: average intra-connectivity within subgraph and average inter-connectivity between subgraphs. The variables included in Equation. \ref{eq:equation9} are explained as follow: $N$ refers to the number of graph partition, $N_i$ is the number of vertices within a subgraph, $coh$ and $cop$ are the degree of cohesion and coupling, $u$ and $v$ are the number of edges within and between subgraphs respectively. A higher value of MQ indicates a better community structure, characterized by higher intra-connectivity and lower inter-connectivity. In the context of microservice candidate extraction, a larger MQ value for the microservice candidate set confirms that the extracted candidates adhere to the software design principle of high cohesion and low coupling.

\begin{equation}
\begin{aligned}
&MQ = \frac{1}{N}\sum_{i = 1}^N \text{coh}_i - \frac{1}{N(N - 1)/2}\sum_{i \ne j}^N \text{cop}_{i,j}, \\
&\text{coh}_i = \frac{u_i}{N_i^2}, \\
&\text{cop}_{i,j} = \frac{\sigma_{i,j}}{2(N_i \times N_j)}.
\end{aligned}
\label{eq:equation9}
\end{equation}

Directly using Eq. \ref{eq:equation9} to calculate MQ for weighted graph structure results in exceeding the defined range, rendering the metric ineffective in its evaluative capacity. We modify it to be suitable for weighted graph calculation and term it as the weighted modularity quality (MQw), presented in Eq. \ref{eq:equation10}. Our modification principle is to preserve the value range of MQ from -1 to 1, while keeping the definition rules of $coh$ and $cop$ unchanged.

\begin{equation}
\begin{aligned}
\begin{alignedat}{2}
&MQ_w = \frac{1}{N}\sum_{i = 1}^N \text{coh}_i' - \frac{1}{N(N - 1)/2}\sum_{i \ne j}^N \text{cop}_{i,j}', \\
&\text{coh}_i' = \frac{u_i'}{N_i^2 + u_i' - u}, \\
&\text{cop}_{i,j}' = \frac{\sigma_{i,j}'}{2(N_i \times N_j) + \sigma_{i,j}' - \sigma_{i,j}}.
\end{alignedat}
\end{aligned}
\label{eq:equation10}
\end{equation}

\begin{table}[tbp]
  \centering
  \renewcommand{\arraystretch}{1.25} 
  \caption{Detail information of legacy systems.}
  \label{tab:my_table}
  \begin{threeparttable}
  \begin{tabular}{|c|c|c|c|}
    \hline
    \textbf{Legacy system} & \textbf{Version} & \textbf{C\#*} & \textbf{LOC**} \\
    \hline
    JPetstore-6 & 6.10 & 24 & 1409 \\
    \hline
    SpringBlog & 1.0 & 46 & 1539 \\
    \hline
    Solo & 4.40 & 139 & 13501 \\
    \hline
  \end{tabular}
  
  \begin{tablenotes}
    \footnotesize
    \item[*] Number of class files.
    \item [**] Lines of code.
  \end{tablenotes}
  
  \end{threeparttable}
  \label{tab2}
\end{table}

\subsection{Dataset and Baseline}
This subsection presents the legacy systems and baselines utilized in our experiments for microservice candidate extraction. Due to the planned application of the method to IIoT's legacy systems and the necessity for experiment reproducibility, we selected open-source Java Web programs from GitHub as the datasets for our microservice candidate extraction experiments. Three open-source programs were selected: JPetstore-6, SpringBlog\footnote{https://github.com/Raysmond/SpringBlog}, and Solo\footnote{https://github.com/88250/solo}. JPetstore-6 is a pet store system, whereas the other two are blog systems. Detailed information about these three legacy systems is presented in Tab. \ref{tab2}. In selecting the experimental baseline methods, we compare our proposed method with the state-of-the-art methods MEM\cite{mazlami2017extraction}, FOSCI\cite{2019Trans}. As an additional baseline approach, we incorporate a method using spectral clustering but solely relies on static analysis tools, named Static.

\begin{figure*}[tbp]
\centerline{\includegraphics[width = 1.0\textwidth]{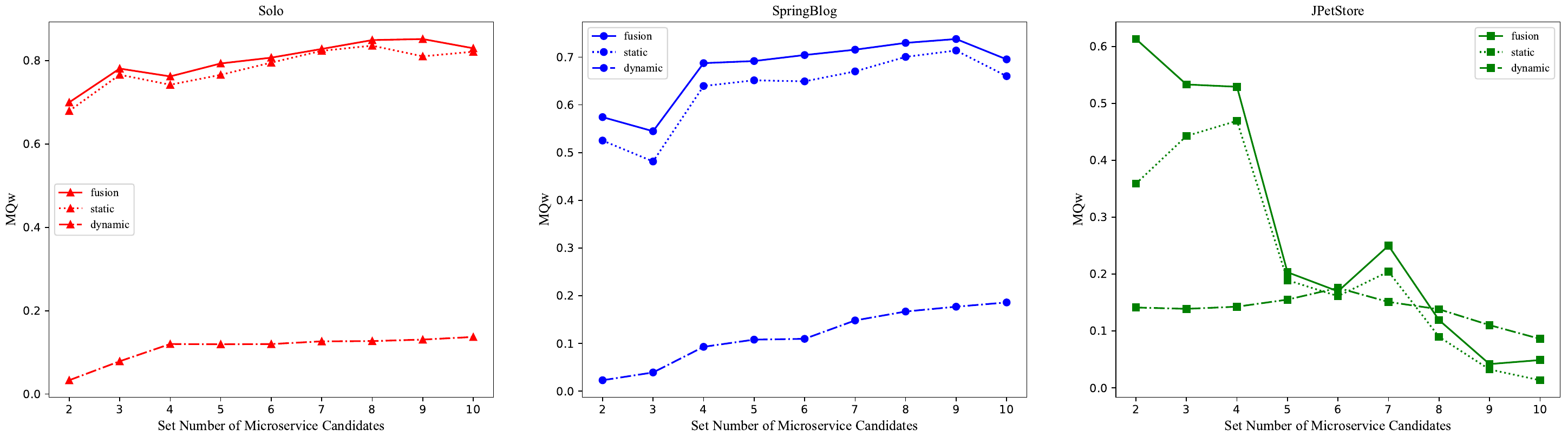}}
\caption{Performance comparison experiments under different analysis tools.}
\label{fig: exp_1}
\end{figure*}

\begin{table*}[htbp]
  \centering
  \renewcommand{\arraystretch}{1.25} 
  \caption{Cohesion and coupling table.}
  \label{tab:comparision table}
  \begin{tabular}{*{13}{c}}
    \toprule
    \multirow{2}{*}{\textbf{Subject}} & \multicolumn{4}{c}{$coh'$} & \multicolumn{4}{c}{$cop'$} & \multicolumn{4}{c}{$MQ_w$}  \\
    \cmidrule(lr){2-5}  \cmidrule(lr){6-9} \cmidrule(lr){10-13}
    & \textbf{Fusion} & \textbf{Static} & \textbf{MEM} & \textbf{FOSCI} & \textbf{Fusion} & \textbf{Static} & \textbf{MEM} & \textbf{FOSCI} & \textbf{Fusion} & \textbf{Static} & \textbf{MEM} & \textbf{FOSCI} \\
    \hline
    \textbf{JPetStore} & 0.992 & 0.991 & 0.215 & 0.211 & 0.379 & 0.633 & 0.115 & 0.039 & 0.613 & 0.358 & 0.1 & 0.172\\
    \textbf{SpringBlog} & 0.682 & 0.65 & 0.195 & 0.535 & 0.003 & 0.004 & 0.092 & 0.153 & 0.679 & 0.646 & 0.103 & 0.382 \\
    \textbf{Solo} & 0.846 & 0.816 & 0.219 & 0.487 & 0.026 & 0.038 & 0.088 & 0.148 & 0.82 & 0.778 & 0.131 & 0.339\\
    \bottomrule
  \end{tabular}
\end{table*}

\subsection{Experiment Result}
For \textit{\textbf{RQ1}}, we perform microservice candidate extraction using both the Fusion and Static methods on the aforementioned three datasets. After carefully considering the sizes of the three datasets, we set the desired number of candidates to be between 2 and 10. Each dataset underwent 100 epochs of experiments for every expected number of candidates. Subsequently, we compute the median of the MQw values obtained from the 100 epochs of experiments and present a line graph for comparison, illustrated in Fig. \ref{fig: exp_1}. The result in this figure shows that the dynamic and static fusion method yields higher MQw values than the method that solely relies on static analysis tools, regardless of the expected number of candidates across the three datasets. This demonstrates that the skill of fuse vertex attributes with edge attributes can indeed lead to improved results in graph partitioning. When using methods that solely rely on dynamic analysis tools to generate graph structures, the resulting candidates exhibit significantly lower modularity quality compared to the previous two analysis strategies. Consequently, relying solely on dynamic analysis strategies for identifying microservice candidates in legacy systems can introduce significant errors. Concurrently, adhering to the software design principle of high cohesion and low coupling, we determine the optimal number of microservice candidates for each dataset as the one with the highest MQw value. It will serve as a reference for setting essential parameters in subsequent experiments.

\begin{figure}[tbp]
\centerline{\includegraphics[width = 0.8\columnwidth]{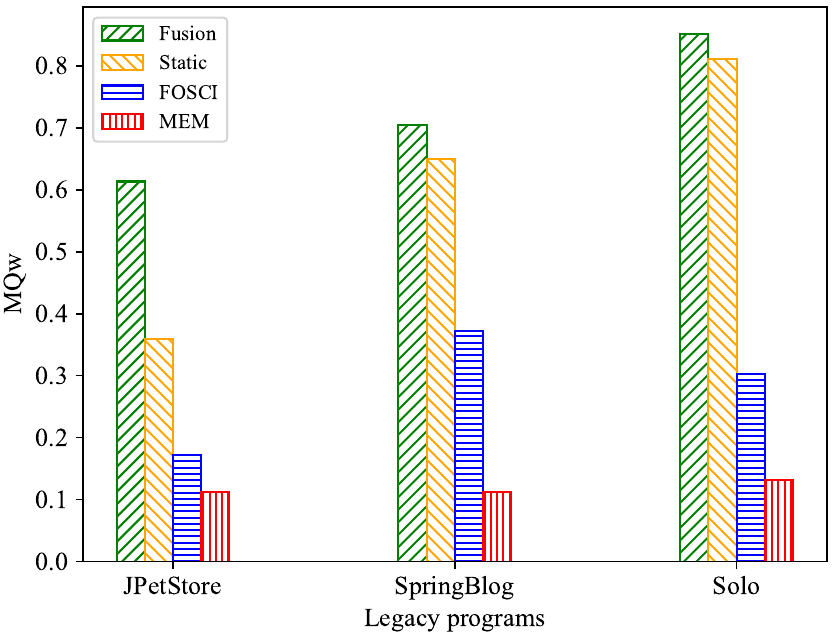}}
\caption{Performance comparison experiments between baseline methods.}
\label{fig: exp_2}
\end{figure}

Next, we address \textit{\textbf{RQ2}} through a comparative experiment. The baselines are shown in the previous subsection. Considering the inadequate performance of the microservice candidate identification method solely relying on dynamic analysis, we excluded this strategy from the experimental comparison. We use MQw as the evaluation metric among the baselines. Our approach for selecting the number of candidates for each baseline in the microservice candidate extraction process is to have each method extract its optimal number of candidates. Based on Fig. \ref{fig: exp_1}, we can determine the optimal number of candidates for the proposed Fusion and Static methods in three datasets: 2, 6, and 9, respectively. The optimal number of candidates for MEM and FOSCI is determined based on the experiment code and supplementary material they provide. Finally, the comparison is conducted by creating a histogram, as illustrated in Fig. \ref{fig: exp_2}. In this figure, we can see the Fusion extraction method, as depicted exhibits significant superiority over the other two baseline methods in terms of the cohesion and coupling metric. This demonstrates that the proposed method is capable of extracting microservice candidates that adhere more closely to the software design principle of high cohesion and low coupling. Conversely, for the other baseline methods, MEM produces sets of microservice candidates with lower modularity quality across all three datasets. FOSCI demonstrates a stronger extraction effect in datasets with a large number of class files, such as SpringBlog and Solo. To further demonstrate the superiority of our proposed method in terms of performance metrics, we calculated the average cohesion $coh'$ and average coupling degree $cop'$ of various microservice extraction methods for each legacy system under the optimal number of candidates, and organized them into Table \ref{tab:comparision table}. We can see that our proposed Fusion method outperforms in the cohesion comparison. The Static method using only static analysis can also achieve high cohesion scores, but is not as good as Fusion on coupling metrics, especially if the legacy system is small. Since MEM is domain-focused, it underperforms on cohesion metrics in all three datasets. As we speculate, the FOSCI approach can be highly cohesive in identifying microservice candidates for large legacy systems, but it cannot achieve optimal performance because it requires comprehensive optimization across multiple objectives. In summary, our proposed Fusion method surpasses other baseline methods in terms of extracting highly modular microservice candidates.

\section{Conclusion}
In this paper, we present an automated method to identify microservice candidates utilizing a graph-optimized clustering algorithm. In contrast to current state-of-the-art approaches, our method is less susceptible to the influence of legacy systems and subjective judgments, allowing for easier implementation. Experimental results showcase the superiority of our method in extracting candidates that adhere to the principles of microservice architecture design. 

In our future work, we aim to enhance our approach by considering multiple design principles in the microservice architecture and incorporating graph neural networks to identify microservice candidates for multi-objective optimization. Additionally, we will employ our proposed method to identify microservice candidates for the legacy industrial software in the actual industrial line. The identified microservice candidates will undergo refactoring and be deployed on a container-based end-edge-cloud interconnection platform for comprehensive functional and performance testing.

\section*{Acknowledgements}
\noindent This work was supported in part by the National Key R\&D Program of China (No. 2021YFB3300100), and the National Natural Science Foundation of China (No. 62171062).

\bibliographystyle{IEEEtran}
\bibliography{reference}

\end{document}